# Big Data Quality: A systematic literature review and future research directions

Mostafa Mirzaie

Department of Computer Engineering, Ferdowsi University of Mashhad (FUM), Iran,
mostafa.mirzaie@mail.um.ac.ir

Behshid Behkamal

Department of Computer Engineering, Ferdowsi University of Mashhad (FUM), Iran, behkamal@um.ac.ir

Samad Paydar

Department of Computer Engineering, Ferdowsi University of Mashhad (FUM), Iran, s-paydar@um.ac.ir

## Abstract

One of the most significant problems of Big Data is to extract knowledge through the huge amount of data. The usefulness of the extracted information depends strongly on data quality. In addition to the importance, data quality has recently been taken into consideration by the big data community and there is not any comprehensive review conducted in this area. Therefore, the main purpose of this study is to provide a systematic literature review for those researchers who are interested in the big data quality subject. Through careful study of the selected papers, we propose a research tree that divides the works based on the type of processing, task, and technique. Further, the challenges and research directions are discussed.

## 1. Introduction

Recent developments in information and communication technology (ICT) have led to mass production of data by social networks, sensor networks, and other Internet-based applications of various domains like healthcare. The vast amount of data that is generated with a high speed from various internet sources is called Big Data [1]. Recently, organizations have concluded that processing big data, especially the data coming from Twitter and Facebook can provide a significant impact on increasing the business's effectiveness and added values [2]–[6]. Figure 1 shows the significant problems caused by the poor quality of big data.

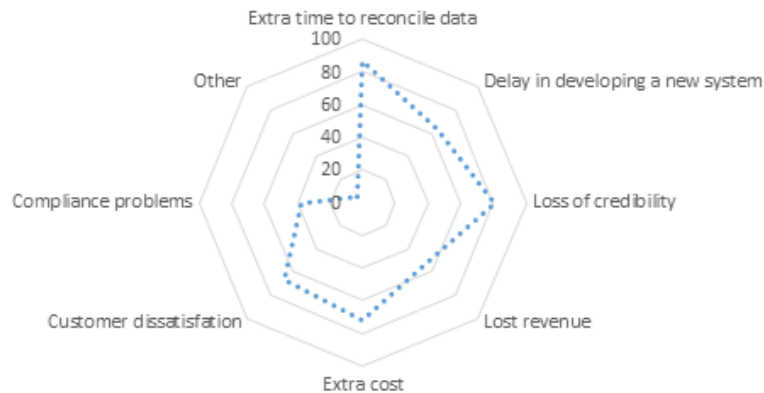

**Figure 1: Problems due to low quality (redrawn based on** [7]**)**

As shown in Figure 1, low data quality may cause problems such as extra cost and allocating more time to reconcile data. The rest of the paper is organized as follows: the search methodology, which includes the planning and conducting phases, is described in Section 2. In Section 3, the research tree obtained from the study of the papers is shown and explained. Studies related to stream processing in Section 4, batch processing in Section 5 and the hybrid methods in Section 6 are described. The review papers are described and compared with the systematic review presented in this paper in Section 7. In Section 8, the results of the systematic review are presented and provide useful information to those researchers who are interested in the big data quality area. Challenges and future work in Section 9, and finally, Section 10 concludes the paper.

## 2. Search Methodology

The methods [8], [9] are employed in order to conduct the presented SLR. In [8] an SLR is performed to review the papers in linked open data, and in [9] an SMS for design patterns is presented to obtain research tree, authors, conferences, and other statistics related to this area. In addition to the research tree and the obtaining of information on various profiles, the papers of the big data quality domain have also been reviewed and compared. Our proposed process has two main steps: 1. Planning and 2. Conducting. Details of our method are discussed below.

### 2.1. Planning Phase

Figure 2 shows the planning process of systematic review.

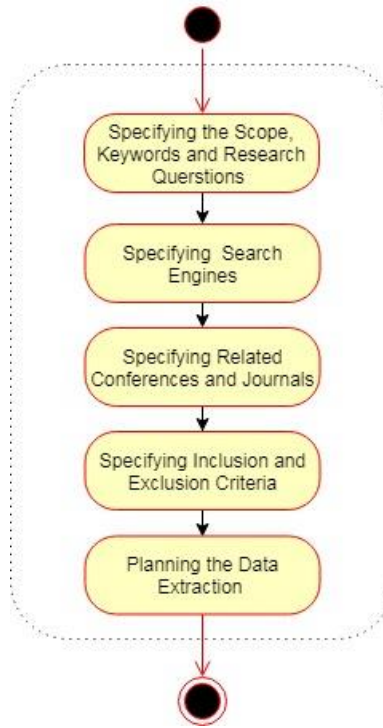

**Figure 2: Planning the Systematic Review**

As shown in Figure 2, this process has five steps: 1) specifying the scope, keywords and research questions, 2) specifying the search engines, 3) specifying related conferences and journals, 4) specifying inclusion and exclusion criteria, and 5) planning the data extraction.

### 2.2. Conducting Phase

After identifying the search strategy and related requirements, they must be applied to the conducting step. The conducting step is divided into three phases.

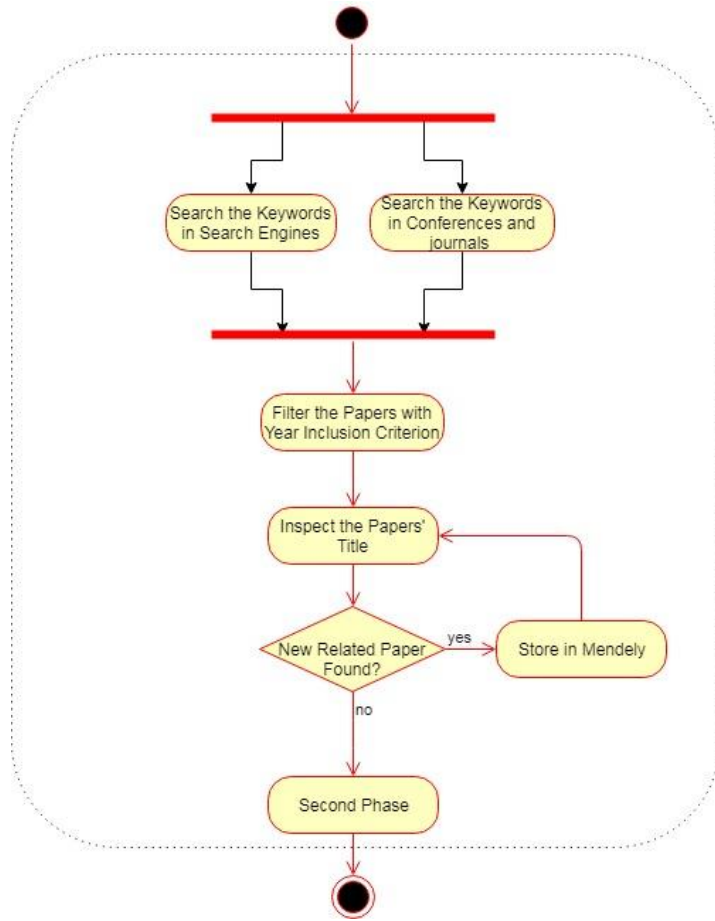

**Figure 3: Conducting the Systematic Review (1st Phase)**

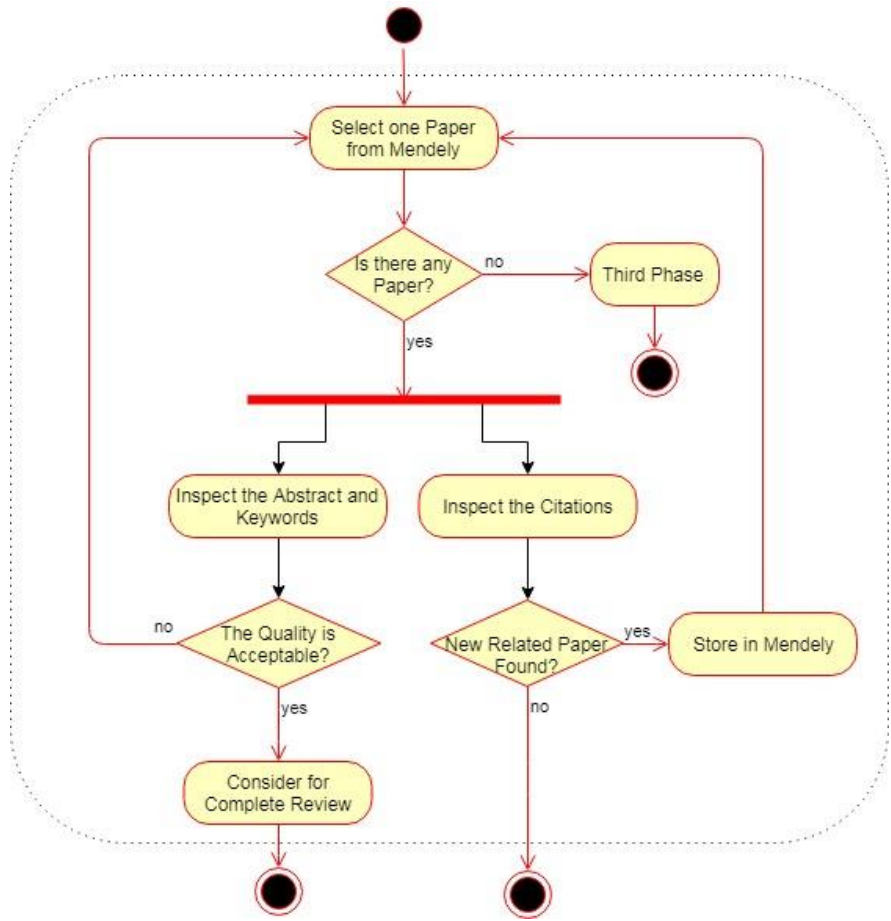

**Figure 4: Conducting the Systematic Review (2nd Phase)**

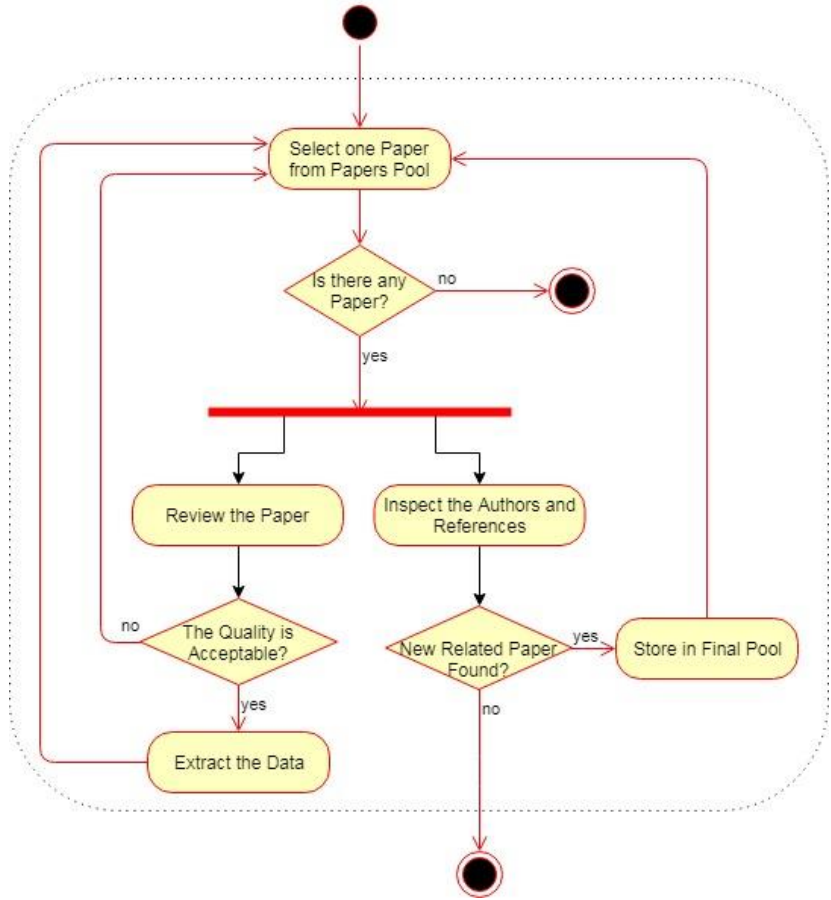

**Figure 5: Conducting the Systematic Review (3rd Phase)**

## 3. Research Tree

After studying all the selected papers, we have analyzed the data that has been extracted, and we have obtained the research tree shown in Figure 6. Through analysis of this research tree, it is possible to answer the first research question, i.e., RQ1. The numbers in each path of the research tree are the number of studies on that path.

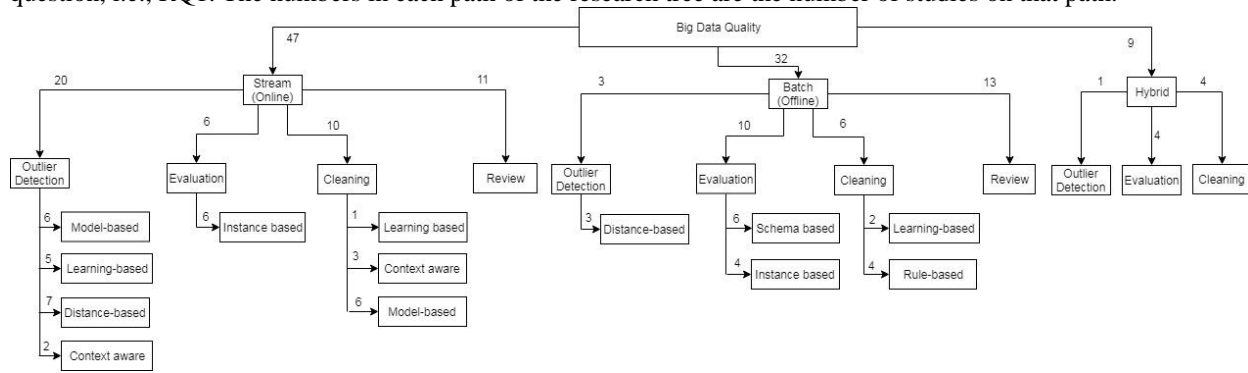

**Figure 6: Research Tree**

## 4. Stream Processing Methods

Stream processing methods are discussed in this section. As described in Figure 7, these methods are divided into four categories: outlier detection, evaluation, cleaning, and review papers.

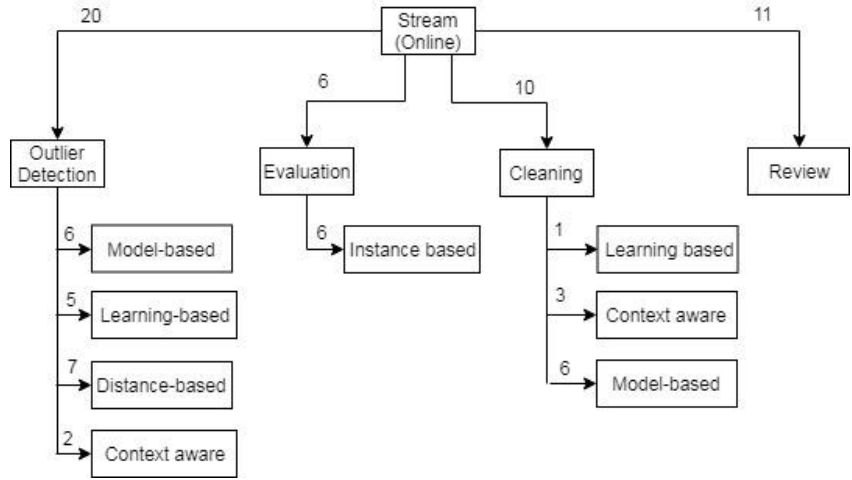

Figure 7: Research Tree of Stream Processing Methods

## 5. Batch Processing Methods

This section describes the papers that use batch processing model to improve the quality of big data. As indicated in Figure 8, based on the primary purpose of the proposed methods, we have divided these works into four categories: outlier detection, evaluation, cleaning, and review papers.

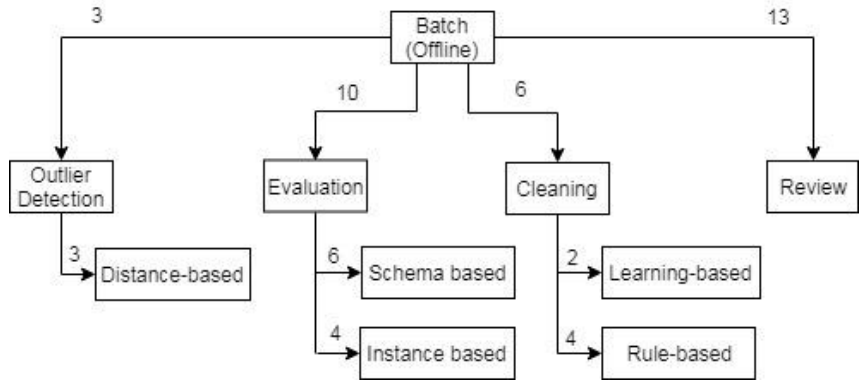

Figure 8: Research Tree of Batch Processing Methods

## 6. Hybrid Methods

Some methods work both on batch and stream data, which usually use static data to build a model and apply the model to the data stream. Sometimes authors run a specific algorithm on both types of data. Because there are no review papers, as shown in Figure 9, these methods are divided into three categories: Outlier Detection, Evaluation, and Cleaning. Since the number of papers in hybrid methods is limited, there is no other classification on each category.

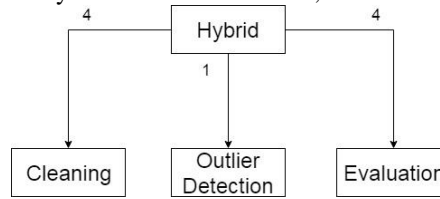

Figure 9: Research Tree of Hybrid Methods

## 7. Comparative Evaluation

In this section, review papers that include Stream [10], [11], [20], [12]–[19] and Batch [21], [22], [31]–[33], [23]–[30] are explained in details and then compared.

## 7.1. Comparison of Review papers

In this section, review papers are compared according to different criteria. **Error! Reference source not found.** compares all review papers mentioned in Sections **Error! Reference source not found.** and **Error! Reference source not found.**, with the proposed systematic review.

## 8. Results

In this section, we refer to the results of the systematic review in order to answer the second research question, i.e. RQ2. Through analysis of the title and keywords of the papers, a tag cloud is produced which is presented in Figure 10.

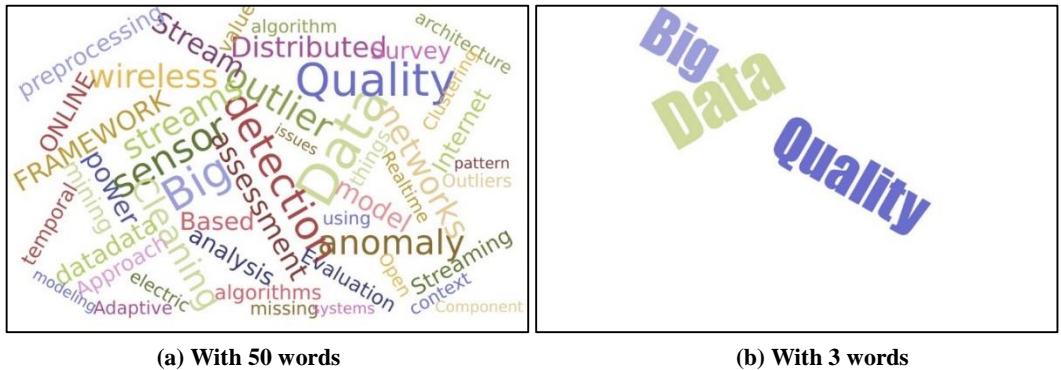

(a) With 50 words          (b) With 3 words

Figure 10: Tag Cloud

Figure 11 illustrates the number of papers on Big Data Quality during the last decade (from 2007 to 2018), based on processing type.

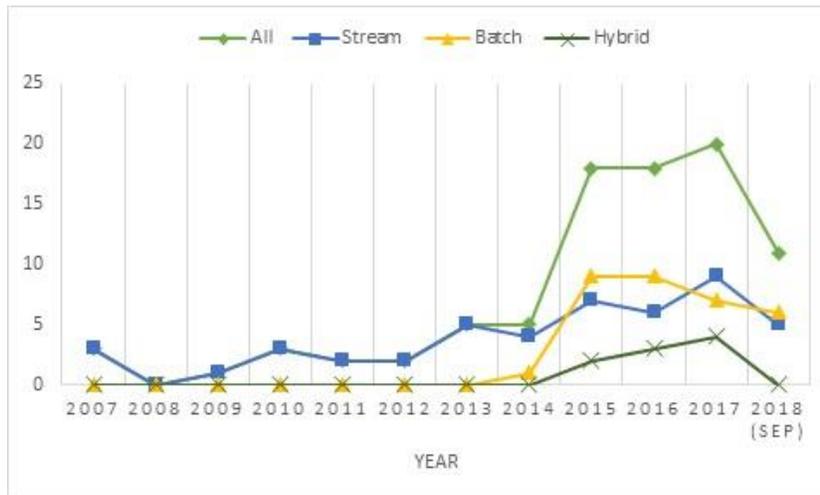

Figure 11: Frequency of publications per year

Additionally, the number of papers in terms of processing type is shown in Figure 12. Due to the expansion of the use of social networks and the increasing use of the Internet of things, the data stream has been used in many papers, and this is illustrated in Figure 12. A smaller number of studies is associated with batch and hybrid processing.

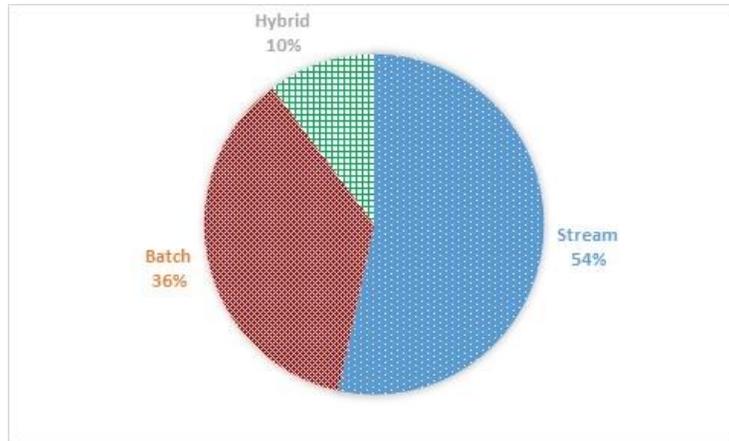

**Figure 12: Number of studies in terms of processing type**

In Figure 13, the number of papers is displayed in terms of both the processing type and the task

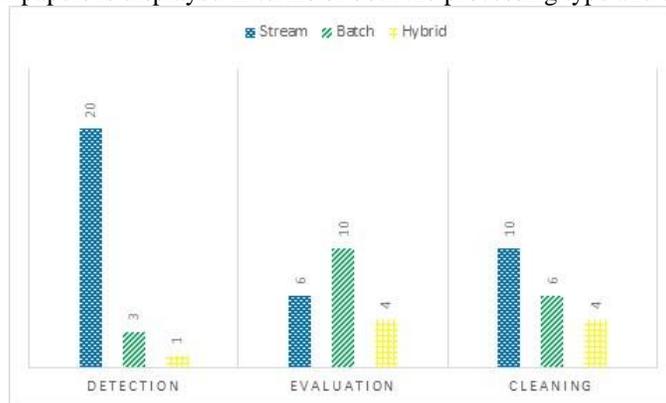

**Figure 13: Number of studies in terms of processing type and task**

The number of papers published in big data quality era is presented in terms of the technique used and the type of processing in Figure 14.

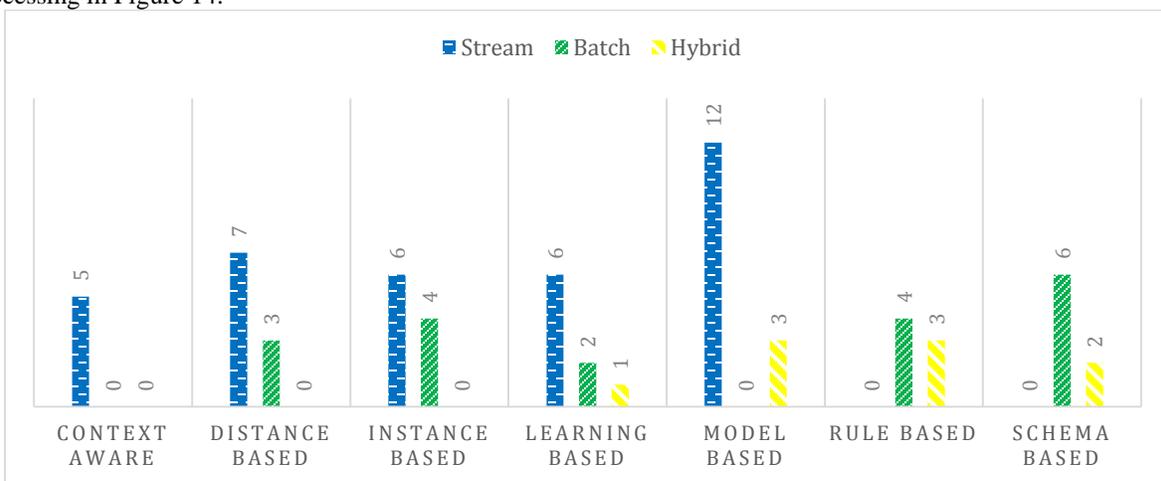

**Figure 14: Number of studies in terms of techniques and processing type**

The distribution of the studies from the point of view of their application domain is shown in Figure 15.

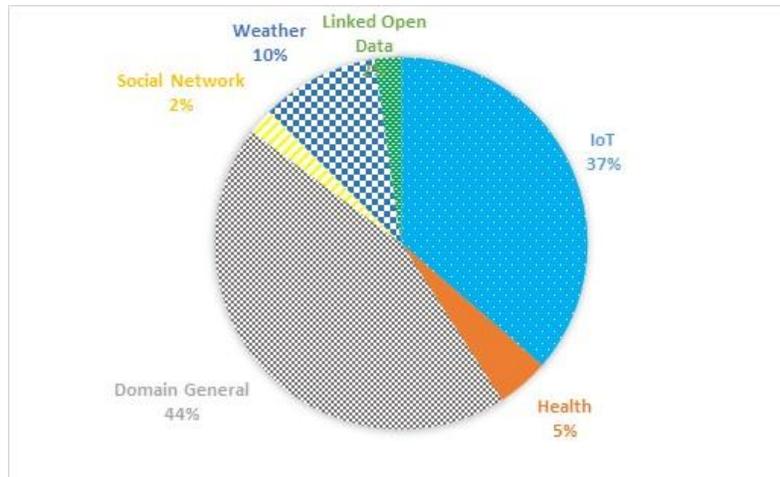

**Figure 15: The domain of studies**

Table 1 refers to the top 5 affiliations that have published studies in big data quality scope.

**Table 1: Top 5 affiliations**

| Affiliation Title | Number of Studies |
|---|---|
| University of Twente | 12 |
| Politecnico di Milano , Concordia University | 7 |
| Northeastern University, Cadi Ayyad University | 6 |

The distribution of the affiliations based on countries with at least four papers is shown in Figure 16.

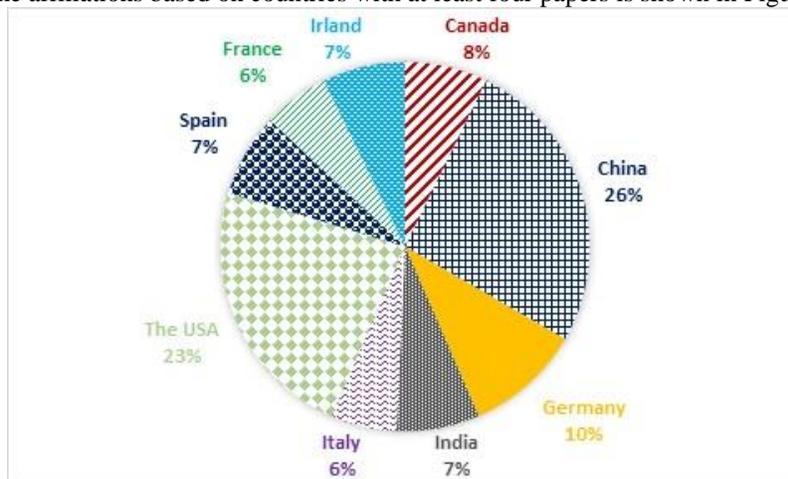

**Figure 16: Active countries in big data quality scope**

The active conferences and journals is also listed in Table 2.

**Table 2: Top conferences and journals with at least two papers**

| Name | Type | Rank* | Number of studies |
|---|---|---|---|
| IEEE International Conference on Big Data | Conference | N/A | 6 |
| IEEE International Congress on Big Data | Conference | N/A | 6 |
| International Conference on Information Quality (ICIQ) | Conference | B1 | 3 |
| Future Generation Computer Systems | Journal | JCR Q1 (IF: 4.639) | 2 |

| Knowledge and Information Systems | Journal | JCR Q1 (IF: 2.247) | 2 |
| Journal of Big Data | Journal | Q1 | 2 |
| Journal of Data and Information Quality | Journal | Q2 | 2 |
| International Journal of Computers and Applications | Journal | Q4 | 2 |
| ACM International Conference on Distributed and Event-based Systems | Conference | N/A | 2 |

* Taken from "http://www.conferenceranks.com" for conferences and "https://www.scimagojr.com" for journals

## 9. Challenges and Future Works

This section, which refers to the challenges in big data quality, is intended to answer the third research question, i.e. RQ3. In general, the challenges and limitations of big data quality can be divided into three categories: source dependent, inherent and technique dependent.

## 10. Conclusion

In this study, a systematic literature review in the big data quality era was conducted to address three research topics including (a) Research topics includes the type of processing, task and the technique used, (b) Active researchers, research institutes, countries and venues, and (c) Challenges and unsolved problems. So, a total of 419 studies were identified, and 170 papers were thoroughly studied, and finally, 88 research papers were included in the final paper pool. Also, the information necessary to answer the specified research questions were extracted from these studies, and the results show that the big data quality is an active and also attractive subject in the last decade.